\documentclass[10pt,aps,twocolumn,preprintnumbers,amsmath,amssymb,footinbib,superscriptaddress,prl
]{revtex4-2}
\usepackage{sistyle}
\usepackage{cancel}
\usepackage[dvipsnames]{xcolor}
\usepackage{graphicx}% Include figure files
\usepackage{dcolumn}% Align table columns on decimal point
\usepackage{bm}% bold math
\usepackage{braket}
\usepackage{hyperref}
\usepackage[all]{hypcap}
\usepackage{easyReview}

\begin{document}

\title{Generation of classical non-Gaussian states by squeezing a thermal state into non-linear motion of levitated optomechanics}

\author{R.~Muffato}
\affiliation{School of Physics and Astronomy, University of Southampton, SO17 1BJ, Southampton, England, UK}
\affiliation{Department of Physics, Pontifical Catholic University of Rio de Janeiro, Brazil}
\author{T.S.~Georgescu}
\affiliation{School of Physics and Astronomy, University of Southampton, SO17 1BJ, Southampton, England, UK}
\author{J.~Homans}
\affiliation{School of Physics and Astronomy, University of Southampton, SO17 1BJ, Southampton, England, UK}
\author{T.~Guerreiro}
\affiliation{Department of Physics, Pontifical Catholic University of Rio de Janeiro, Brazil}
\author{Q.~Wu}
\affiliation{Centre for Quantum Materials and Technologies, School of Mathematics and Physics, Queen's University Belfast, BT7 1NN, United Kingdom}
\author{D.A.~Chisholm}
\affiliation{Centre for Quantum Materials and Technologies, School of Mathematics and Physics, Queen's University Belfast, BT7 1NN, United Kingdom}
\author{M.~Carlesso}
\affiliation{Department of Physics, University of Trieste, Strada Costiera 11, 34151 Trieste, Italy}
\affiliation{Istituto Nazionale di Fisica Nucleare, Trieste Section, Via Valerio 2, 34127 Trieste, Italy}
\affiliation{Centre for Quantum Materials and Technologies, School of Mathematics and Physics, Queen's University Belfast, BT7 1NN, United Kingdom}
\author{M.~Paternostro}
\affiliation{Universit\`a degli Studi di Palermo, Dipartimento di Fisica e Chimica - Emilio Segr\`e, via Archirafi 36, I-90123 Palermo, Italy}
\affiliation{Centre for Quantum Materials and Technologies, School of Mathematics and Physics, Queen's University Belfast, BT7 1NN, United Kingdom}
\author{H.~Ulbricht}
\email[Correspondence email address: ]{H.Ulbricht@soton.ac.uk;}
\affiliation{School of Physics and Astronomy, University of Southampton, SO17 1BJ, Southampton, England, UK}

\date{\today}

\begin{abstract}
\noindent
We report on an experiment achieving the  dynamical generation of non-Gaussian states of motion of a levitated optomechanical system. We access intrinsic Duffing-like non-linearities by thermal squeezing an oscillator's state of motion through rapidly switching the frequency of its trap. We characterize the experimental non-Gaussian state against expectations from simulations and give prospects for the emergence of genuine non-classical features.
\end{abstract}

\maketitle
\noindent
\emph{Introduction} -- Macroscopic quantum states are regarded as ideal platforms to test for the existence of intrinsic limitations in quantum formalism~\cite{arndt2014testing} and the effects of gravity in quantum systems~\cite{carney2019tabletop}. While various platforms have been considered for generating such states~\cite{nimmrichter2013macroscopicity}, levitated systems have emerged as a front-runner in light of the possibility to achieve remarkable environmental isolation~\cite{gonzalez2021levitodynamics} and embed control techniques in their dynamics. Engineering non-classical states of levitated mechanical systems is achieved by inducing non-linear dynamics to sufficiently coherent initial states of motion~\cite{rodamacroscopic, rakhubovsky2019nonclassical}. Accessing non-linearity by motional squeezing -- sometimes assisted by a dark trap --  can be instrumental to preparing non-classical states~\cite{bonvin2023state}. The fundamental building block of such operations, namely the squeezing of thermal states of motion of levitated particles, has been demonstrated by rapid (i.e. faster than the oscillation frequency) switching between different trap frequencies~\cite{rashid2016experimental}, as theoretically proposed by Janszky et al.~\cite{janszky1986squeezing}, or using cavities~\cite{penny2023sympathetic, li2023mechanical}. Such switches can be treated as pulses applied to the particle's spring constant and have been used to drive a classical harmonic oscillator~\cite{arnol2013mathematical}. Repeated gentle squeezing pulses applied to the thermal motional state's momentum can increase the particle's oscillation amplitude, allowing it to explore the non-linear flat extremities of the trapping potential while remaining trapped.
Correspondingly, the dynamics can pick up Duffing-like non-linearities, as already demonstrated~\cite{setter2019characterization}. This scheme dynamically shapes the potential landscape creating bi-stability. Classical effects have been discussed before, applied in fields such as memory elements \cite{frimmer2019rapid}, signal amplification via stochastic resonance \cite{ricci2017optically}, non-equilibrium physics \cite{gieseler2014dynamic}, nonlinear dynamics, synchronization \cite{gieseler2014nonlinear}, and active escape dynamics \cite{militaru2021escape}. It is also predicted to enhance force sensing capabilities of levitated mechanics~\cite{cosco2021enhanced}. \\
In this paper, we experimentally demonstrate how a sequence of repeated pulses can be used to dynamically produce a significantly non-Gaussian  state of motion by accessing intrinsic trapping-potential non-linearities and controlling their effects. We show the time-trace, in phase space, of the motional state resulting from the application of our protocol to an initial thermal state, demonstrating the emergence of non-trivial transient bimodal distributions whose relaxation dynamics we characterize experimentally. Moreover, we provide a quantitative perspective for the emergence of genuine quantum features of motion from a sufficiently coherent initial state of the particle~\cite{tebbenjohanns2021quantum, delic2020cooling, magrini2021real, kamba2022optical, ranfagni2022two}. We show how this would result in negative phase-space distributions, thus unambiguously demonstrating non-classicality, a key feature for quantum information processing and fundamental tests of quantum mechanics.

\begin{figure*}
\includegraphics{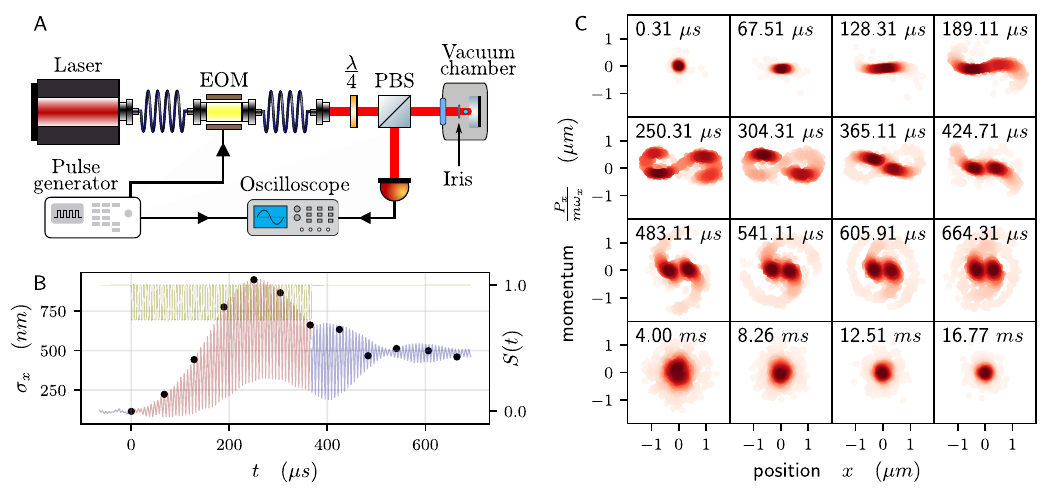}
\caption{\label{fig:wide} Panel A: Schematic of experimental setup. A silica nanoparticle is
trapped by an optical tweezers in vacuum. An electro-optical modulator (EOM) controlled by a pulse generator modulates the power of the laser. The oscilloscope receives the signal from the particle. Its acquisition is triggered by a synchronous signal from a pulse generator. Panel B: Experimental time evolution of the standard deviation of the particle position as measured by the photodiode (Blue/red line). Green line is the intensity modulation function $S(t)$, see Eq.(~\ref{Langevin_1D}), applied to the trapping laser beam via the EOM. Panel C: Experimental time-trace of the phase-space distribution of an initial thermal state of motion of the particle subjected to our protocol. The first 12 snapshots of distribution are indicated as black dots in panel B. The last line depicts the thermalization occurring at a longer time scale.}
\end{figure*}

\begin{figure}[b!]
\centering
\includegraphics[scale=1]{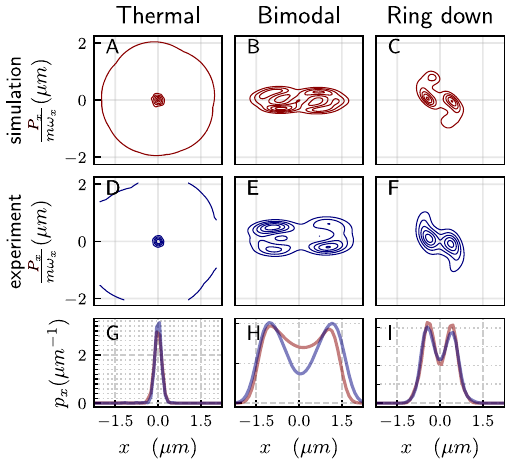}
\caption{\label{fig:std_seq2} Comparisons between simulations (red lines) and experimental reconstructions (blue lines) of the phase-space distributions associated with paradigmatic states of motion of the levitated particle. Panels {A} and D: Initial thermal state; Panels B and E: initial non-Gaussian bimodal state; Panel C and F: The initial distribution is that of the motional state achieved by stopping the dynamical protocol. Such state will eventually relax back to a thermal configuration. Panels G-I show the projection of the phase-space distribution, with $p_{x}$ the position probability function for the three initial states described above.} 
\end{figure}

\noindent
\emph{Theoretical model} -- We refer to the setting illustrated in Fig.~\ref{fig:wide}A, which depicts a nanoparticle trapped in an optical tweezer in vacuum. In general, the particle motion occurs  in all three spatial directions ($x, y, z$), including cross coupling between orthogonal directions of motion \cite{gieseler2014nonlinear} and possibly involving rotational motion as well. However, our analysis is focused only at the $x$ motion.
We model the stochastically driven and damped oscillatory motion of a levitated particle in a vacuum by a one-dimensional Langevin-type equation for the time-dependent position $x(t)$ of the centre-of-mass motion of the particle of mass $m$ in the trap
\begin{equation}
    \ddot{x}(t) = -\Gamma_m \dot{x}(t) + S(t)\frac{F(x)}{m} + \frac{\mathcal{F}_{fluct}(t)}{m}, 
    \label{Langevin_1D}
\end{equation}
where, in the regime dominated by the background gas pressure $P$, the damping rate can be written as~\cite{gieseler2013thermal}
\begin{equation}
    \Gamma_m = \frac{64r^2P}{m\overline{v}_{gas}}, 
    \label{gas_damping_Gieseler}
\end{equation}
with $\overline{v}_{gas}$ the root mean squqre thermal velocity of the gas molecules at room temperature, while $r$ and $m$ is the radius and mass of the trapped silica particle. The second term in the right-hand side of Eq.~\eqref{Langevin_1D} is the position-dependent restoring force originating from the interaction with a Gaussian light beam in the Rayleigh regime. In the large oscillation-amplitude limit, when the non-linearity becomes effective, this force is given by
\begin{equation}
    F(x) = \frac{2\pi n_{m}r^3}{c} \left(\frac{n_{r}^2-1}{n_{r}^2+2}\right)\frac{\partial I}{\partial x} ,
    \label{gradient_force}
\end{equation}
with $n_{m}$ and $n_{p}$ being the refractive index of the medium and of the silica particle (here assumed to take the nominal values of 1 and 1.44, respectively), $n_{r}=n_{p}/n_{m}$ the relative refractive index, $c$ the speed of light,  and $I$ the intensity of the trapping laser beam. Eq.~(\ref{gradient_force}) gives rise to a nonlinear force directed along the gradient of the optical potential $U(x)$ such that $F(x)=-\partial_x U(x)$~\cite{jones2015optical,ashkin1986observation,harada1996radiation}, which we control by modulating in time the power of the trapping laser according to the function $S(t)$ in Eq.~(\ref{Langevin_1D}) and as illustrated in Fig.~\ref{fig:wide}B~\cite{rashid2016experimental}. The last term in the right-hand side of Eq.~(\ref{Langevin_1D}) describes the driving of the harmonic motion by random background gas collisions, which are modelled as the time-dependent stochastic force
$\mathcal{F}_{fluct}(t)=\sqrt{2\Gamma_{m}k_{B}T}\eta(t)$. Here, $k_B$ is the Boltzmann's constant and $T$ the temperature of the gas surrounding the oscillating particle~\cite{kubo1966fluctuation}, while $\eta(t)$ is a zero-mean, delta-correlated Gaussian white noise such that $\langle \eta (t) \eta (t') \rangle = \delta(t - t')$. We have integrated Eq.~(\ref{Langevin_1D}) numerically using the Euler-Maruyama method~\cite{volpe2013simulation} and sampling the initial values of position and momentum of the particle from a thermal state. The results of our analysis are shown as the phase-space probability distributions in Fig.~\ref{fig:std_seq2}A-C showing that our one-dimensional description of the dynamics is sufficient to produce the main features of experimental data. 

\noindent
\emph{Experiment} -- The experimental setup is illustrated in Fig.~\ref{fig:wide}A. We use a parabolic mirror to focus light at the wavelength of 1550 nm to a diffraction-limited spot of the diameter of $~1\mu$m. A pulse generator (Berkeley BNC 525) and an electro-optical modulator (EOM, Jenoptik AM1550b) are used to modulate the laser intensity as a square-wave with high and low levels of~\SI{80}{mW} and \SI{57}{mW} respectively, to trap a silica particle of close to spherical shape (\SI{460}{nm} diameter) from a spray in the focal point. We monitor the centre-of-mass motion of the trapped particle by detecting a small fraction of Rayleigh back-scattered light from the trapping field with a single photodiode detector (see Ref.~\cite{vovrosh2017parametric} for technical details). An iris is mounted on top of the parabola to suppress unwanted light reflected from the flat edges of the mirror and make sure that all the detected signal comes from the particle only. This guarantees the reliable assessment of the particle's center-of-mass position with only negligible distortions by the bandpass filtering process, as analysed in more detail in the Supplemental Material (SM) Sec.~A. 
Calibration of motional amplitudes and the mass of the trapped particle was done assuming perfect equilibrium of the particle with the background gas at \SI{5}{mbar} by fitting a Lorentzian to the trap frequency peak~\cite{vovrosh2017parametric}. \\
To prepare bimodal states of motion the pressure is further reduced to \SI{1e-2}{mbar} and the particle resonance frequency {$\omega/2\pi=\SI{77}{kHz}$} is evaluated from the spectrum of the signal at the high-power level. Next, thermal squeezing is achieved modulating the scale of the restoring force term of Eq.~\eqref{Langevin_1D} as $S(t)$ with

\begin{equation}\label{equ:control_function_OP}
    {S}(t) = 
    \begin{cases}
    1, &\text{for}~t'=0~\text{or}~t'\in[\tau_{low}, \tau_{low} + \tau_{high}],\\
    0.71, &\text{for}~ t'\in[0, \tau_{low}].
    \end{cases}
\end{equation}
Here $t' = t\!\mod{(\tau_{low}+\tau_{high})}$. Each pulse is timed so that the particle completes a quarter of an oscillation at the high-power level,  $\tau_{high}=\pi/2\omega = \SI{3.25}{\mu s}$, and another quarter of an oscillation at the low-power level, $\tau_{low}=\pi/2\omega\sqrt{S} =\SI{3.48}{\mu s}$. The duration at the low-power level is slightly longer due to the re-scaling of dynamics by the relative decrease of power. One pulse sequence consists of a train of $55$ pulses as constructed on the pulse generator to prepare the bimodal state. Data acquisition by the oscilloscope was synchronised with the pulse generator and $689$ pulse sequences are acquired with $\SI{518}{ms}$ between successive sequences which is about $25$ times longer than the extracted $\SI{20}{ms}$ relaxation time, in good agreement with Eq.(\ref{gas_damping_Gieseler}), allowing the particle to relax back to the thermal state between consecutive runs. We experimentally characterize a Duffing non-linearity $\xi=-0.1\, \mu$m$^{-2}$ for our parabolic mirror trapping potential, $U(x)=\omega ^2 x^2 +w^2\xi x^4$, as described in more detail in Ref.~\cite{setter2019characterization}.

\noindent
\emph{Results and Discussion} -- The trap non-linearity generating the non-Gaussian state is small in typical experimental conditions~\cite{gieseler2013thermal}. Taylor expansion of the gradient force up to third order shows that the inverse of the Duffing parameter is proportional to the square of the beam's waist. Considering it as a measure for the range of  nonlinearity, we note that it is larger than the spread of the thermal state $\sigma_{thermal}=\sqrt{k_{B}T/k}$, where $k$ is the optomechanical spring constant and $T$ the motional temperature. This means that the particle hardly visits the nonlinear part of the potential. By squashing the motional state, we induce sufficient elongation such that the vertexes of the state are eventually extended to nonlinear parts of the potential well and the non-linear effect dominates the dynamics. In these regions, the particle's dynamics is slower due to a softer effective spring constant making the tip of the ellipse lag behind compared to the faster harmonic behaviour near the centre. In the course of sufficiently many weak pulses, the state changes its shape, steadily transforming it from the initial circular thermal Gaussian to a squashed ellipsoidal and eventually to curved spiral where probability mass starts to accumulate, forming a multi-modal state. We emphasize that in general at very large motions a self-homodyne detection might become substantially nonlinear, potentially affecting the phase-space distributions. Possibilities to avoid these effects is to cool the motion to ensure the homodyne detection stays linear~\cite{frimmer2019rapid} and to benchmark homodyne detection against heterodyne measurements to rule out non-linearities in the detection~\cite{dawson2019spectral}.\\
To quantify the non-Gaussianity of the resulting simulated and experimentally obtained states, we adopt a non-dimensional measure of bi-modality \cite{ashman1994detecting}, 
\begin{equation}
A_{D} = \frac{\sqrt2}{\sigma_{T}}|\mu_{1}-\mu_{2}|, 
\label{eq:bimodal}
\end{equation}
where $\mu_{1,2}$ are the positions of the peaks of the distribution and $\sigma_T=\sqrt{\sum^2_{i=1}\sigma^2_i}$ is defined in terms of the individual peak's variances $\sigma_{1,2}$ [cf. Fig.~\ref{fig:std_seq2}H]. By curve fitting a double Gaussian function to our experimental and simulated data we find the values of $A_{D}^{exp} = 3.95$ and $A_{D}^{sim} = 3.32$ in good agreement with 12$\%$ deviation which we identify to be originated by a small difference in the shape of the real trapping potential from the simulated one. The influence of the potential's non-linearity is understood to be negligible in the linear section near the trap centre and to become dominating towards the asymptotically flat edges of the optical trap. Despite knowing general features of the non-linearity of the potential well, its exact profile is not experimentally known. As such, simulations assuming an ideal sharp beam profile were found to deviate more from the experimental results. We also find that the experimental thermal state has some outlier points that overextend the tails of Gaussian distribution, which are not included in the simulations and contribute to a softening of the potential and a larger value for $A^{exp}_D$.  We explain outliers by a build-up of amplitude as a net effect from multiple collisions by background gas particles or other perturbations affecting the trap. %See supplement for more details. 

\noindent
\emph{Perspectives for quantum experiment --}
To investigate nonlinear squeezing effects provided by the protocol for a quantum system, we consider the system Hamiltonian
\begin{align}\label{equ:system_hamiltonian_OP}
    \hat H_s(t) = \frac{\hat p^2_x}{2m} - {S}(t)U_{0} e^{- 2\hat x^2/w^2_{0}} ,
\end{align}
where $\hat x$ and $\hat p_x$ are the position and momentum operators. The potential of the system, represented by the second term, is an inverted Gaussian as that used in the classical simulation with its gradient corresponding to Eq.~(\ref{gradient_force}). Here, $U_0$ sets the energy scale (the depth) of the potential and $S(t)$ is the control given by Eq.~\eqref{equ:control_function_OP}. We assume that the experiment is performed in a low-pressure, cryogenic environment, such that it is possible to perform scattering-free experiment runs \cite{Bykov_2019}. In this regime, the dynamics of the system follows the von-Neumann master equation,
$\dot{\hat \rho}(t) = -\frac{i}{\hbar} [\hat H_s (t) , \hat \rho(t)] - \Lambda[\hat{x},[\hat{x},\hat{\rho}]]$, where the second term describes the decoherence due to photon recoils \cite{Weiss_2021}.\\
For a typical opto-levitating experiment, the size of the quantum system is way smaller than that of the trapping potential. Therefore, the challenge of this quantum experiment is to expand the quantum system to the spread (FWHM) of the potential approximated by the focus beam waist $\texttt{w}_{0}$ before complete decoherence, so that systems like the one reported above can exhibit non-classical features such as negativity in the Wigner distribution $N(t)$. The negativity $N(t)$ is defined by integrating the Wigner distribution over the region of the phase space $\Sigma_-$ where it attains negative values, that is
\begin{equation}\label{equ:wigner_negativity_OP}
    {N}(t) =\iint_{\Sigma_-} \text{d}x\,\text{d}p_x\,{W}_{x, p_x}(t) .%\vert_{{\cal W}<0}.
\end{equation}
For a highly focused Gaussian beam, the beam waist can be reduced to $\omega_0\approx 750$\,nm (the Duffing parameter $\xi\approx -0.3$\,$\mu$m$^{-2}$). We assume that the quantum system can be expanded by a factor of $\zeta=100$ through a fast squeezing protocol \cite{Weiss_2021}. This means that we need to prepare an initial quantum system of size $\Delta x_0\approx 7.5$\,nm. This is difficult, as the zero-point fluctuation of the motion is approximately $\Delta x_{zpf}=\sqrt{\hbar/2 m \omega}=0.01$\,nm for a nanoparticle of $R=50$\,nm ($\rho=2200$\,kg/m$^3$) trapped at frequency $\omega/2\pi=50$\,kHz. 
Indeed, this bottlenecks all the schemes to prepare large mass non-classical states -- the mismatch between the scale of the beam trap and the coherence length of a massive quantum system.
\begin{figure}[b!]
\centering
\includegraphics[width=1\linewidth]{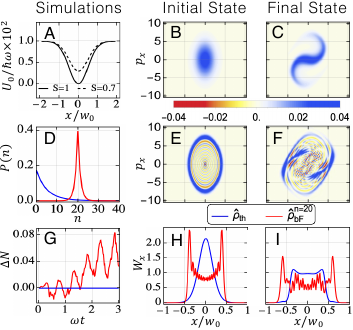}
\caption{\label{fig:quantum_simulation} Simulations of the nonlinear protocol for quantum systems. Panel~A shows the oscillating potentials with $U_0/\hbar\omega =100$ energy levels. Panels~B and E are the Wigner distribution for the initial thermal state $\hat{\rho}_\text{th}$ and the initial blurred Fock state $\hat{\rho}_\text{bF}^{n=20}$, and Panels~C and F are those of the final state at $\omega t = 11/4$. Panel~D shows the population distribution $P(n)$ for the initial states. Panel~G shows the increment of negative Wigner distribution $\Delta N$ over $3$ oscillations. Panel~H and I shows the marginal Wigner distribution $W_x$ of the initial and final systems. Here the blue line represents the thermal state $\hat{\rho}_\text{th}$ and the red line represents the blurred Fock state $\hat{\rho}^{n=20}_\text{bF}$. } 
\end{figure}\\
One way to approach this issue is by considering a smaller nanoparticle and preparing the $n$-th Fock state as the initial state. Indeed, a smaller particle yields a smaller decoherence rate and a larger quantum state. Consider a nanoparticle of $R=4$\,nm trapped at frequency $\omega/2\pi = 80$\,kHz. The decoherence constant $\Lambda$ depends on the particle size (and other experimental details), and we estimate it with the equation \cite{Gonzalez-Ballestero_2019,Seberson_2020}
\begin{equation}
    \Lambda=\frac{7 \pi \varepsilon_0}{30\hbar} \left(\frac{\epsilon_c V E_0}{2\pi}\right)^2 k_0^5,
\end{equation}
where $\varepsilon_0$ is the vacuum permittivity, $\epsilon_c=3{(\epsilon-1)}/{(\epsilon+2)}$ with $\epsilon$ being the relative dielectric constant of the nanoparticle. $V$ is the volume of the nanoparticle, $k_0=2\pi/\lambda$, and $E_0=\sqrt{{4P_0}/{\pi\varepsilon_0\,c\, \texttt{w}_0^2A_xA_y}}$ where $P_0$ is the tweezer power, $\texttt{w}_0$ is the beam waist, $A_x$ and $A_y$ are the asymmetry factor and $c$ is the speed of light. Taking the values in Ref.~\footnote{We take the particle radius $R=4$\,nm, the laser wavelength $\lambda=1550$\,nm, the laser power $P_0=0.5$\,W, the trap width $w_0=750$\,nm, the vacuum permittivity $\varepsilon_0=8.9\times10^{-21}$\,F/nm, the relative dielectric constant $\epsilon=2$ and the asymmetry factors $A_x=1$, $A_y=0.9$.}, the decoherence constant $\Lambda$ is estimated to be $\Lambda \approx 3.28\times10^{19}$\,Hz$\cdot$m$^{-2}$. The corresponding decoherence rate $\Gamma = \Lambda \Delta x_{zpf}^2\approx 5.8$\,Hz, such that $\Gamma/\omega \approx 1.16\times 10^{-5}$.
On the other hand, the $n$-th Fock state has the standard position variance $\Delta x^{n}=\sqrt{(2n+1)\hbar/2m\omega}$, such that $\Delta x^{100}\approx 6$\,nm for the state $\ket{n=100}$. Based on the estimation from Sec.~G in SM, the initial system $\ket{n=100}$ can be expanded by a factor of $\zeta=100$ and reach $\Delta x^{100}_\text{expanded}\approx 600$\,nm with remaining system purity ${\cal P}\approx 0.9$.
\\
We perform simulations with a smaller quantum system to demonstrate the effects of the protocol (details reported in Sec.~H in SM) and show the results in Fig.~\ref{fig:quantum_simulation}. The system considered here has $U_0/\hbar\omega=100$ energy levels in the potential, as shown by Fig.~\ref{fig:quantum_simulation}A. 
We take two kinds of initial state with roughly the same energy -- the thermal state $\hat{\rho}_\text{th}$~\footnote{We take mean excitation number $\braket{n}=(\exp[\hbar\omega/k_BT]-1)^{-1}=4.52$. Here $\omega$ is the oscillation frequency approximated at the low-energy section of our inverted Gaussian system (\textit{e.g.}~in our example $\omega/2\pi =80$\,kHz). The variance of this thermal state is half of that of the blurred Fock state, as shown by Fig.~\ref{fig:quantum_simulation}H.} and the blurred Fock state $\hat{\rho}_\text{bF}^{n=20}$ (which is the Gaussian mixture centered at the Fock state $\ket{n=20}$, defined by Eq.~(S11) in SM) as an example of the aforementioned quantum experiment. Their population distribution $P(n)$ is shown by Fig.~\ref{fig:quantum_simulation}D and their Wigner distribution is shown by Fig.~\ref{fig:quantum_simulation}B and \ref{fig:quantum_simulation}E. We simulate the nonlinear protocol with these two initial states for oscillation time $\omega t=11/4$ and the decoherence rate $\Gamma/\omega = 10^{-5}$, and show the Wigner distribution for the final states in Fig.~\ref{fig:quantum_simulation}C and \ref{fig:quantum_simulation}F. The marginal Wigner distributions $W_x$ for the initial and final states are shown by Fig.~\ref{fig:quantum_simulation}H and \ref{fig:quantum_simulation}I, and the increment of the Wigner negativity $\Delta N = N(t) - N(0)$ is recorded in Fig.~\ref{fig:quantum_simulation}G. 

From the outcomes, we draw the following conclusions. Firstly, taking the thermal state $\hat{\rho}_\text{th}$ as the initial state, we retrieve a similar behavior as in the classical experiment (Gaussian to bimodal distribution) although working at a much lower energy. However, due to the mixture of the Fock states, we observe no increment of negative Wigner distribution when taking this initial state. 

Secondly, taking the mixed Fock state $\hat{\rho}^{n=20}_\text{bF}$ as the initial state (see SM, Sec.~I for other initial states), we still see the nonlinear effect of the protocol (in terms of the increasing separation of two peaks in the marginal Wigner distribution) as well as the increment of the Wigner negativity. 
This protocol shows the potential of generating non-Gaussian states with non-classical features (Figs.~3G-I) by coherently expanding the initial quantum state enough to access the nonlinearity of the potential. 

However, generating such Fock states at current experimental conditions is challenging. The requirements can be potentially relaxed in a more sophisticated experiment. Introducing extrinsic nonlinearity \cite{Hempston_2017}, that can reduce the requirement on expansion rate $\zeta$ through enhancing the Duffing parameter $\xi$. For example, actuating on the system, either externally \cite{Kremer_2024} or parametrically, conditioned on real-time position measurement has the potential to modify the effective dynamics. Another measure is reducing the photon recoil rate $\Gamma$ by taking zero-information measurement \cite{Szigeti_2014} with implementation ideas for levitated systems \cite{gajewski2024backaction,weiser2024back}).  Such improvements could potentially facilitate the initiation of this protocol from a low-energy thermal state, enabling the generation of a bimodal distribution while likely preserving quantum coherence.

Finally, we note that, optomechanical state tomography has been successfully demonstrated in the classical regime for clamped  \cite{vanner2013cooling} and levitated \cite{rashid2017wigner} systems, with several proposals extending the technique to the quantum regime \cite{vanner2015towards,shahandeh2019optomechanical,warszawski2019tomography}, including approaches utilizing neural networks \cite{weiss2019quantum}. However, a specific quantum procedure to extract Wigner negativity for levitated optomechanics systems has yet to be fully developed and requires further research.

\noindent
\emph{Conclusion} -- We have shown experimentally that the controlled access of intrinsic trap non-linearities can be used for the generation of non-Gaussian classical states. 
We show theoretically that, if the motional state is initially prepared to a Fock state, taking fast squeezing protocol allows the system to access the intrinsic trap non-linearities with high remaining purity. 
Experimental improvements can be realized to achieve non-Gaussian states in smaller length scales, thus diminishing decoherence effects. For instance, using stronger non-linearities, such as those from engineered potentials~\cite{rakhubovsky2019nonclassical} and deploying quantum control methods of thermodynamic inspiration~\cite{Qiongyuan_2023}, which will be the focus of forthcoming work. Then, our scheme can be used to generate bimodal states with a genuinely non-classical character, as witnessed by significantly negative Wigner distributions. We remark the experimental realization of a very similar protocol for squeezing in nonlinear mechanical quantum system in the GHz regime with the result of negativity~\cite{marti2024quantum}.
Periodically driving mechanical motion as in our squeezing protocol is also predicted to enhance force sensing capabilities of levitated mechanics~\cite{cosco2021enhanced}. 

\noindent
\emph{Acknowledgements --}
We thank for discussions Jakub Wardak, Elliot Simcox, and Chris Timberlake. We acknowledge funding from the EPSRC International Quantum
Technologies Network Grant {\it Levinet} (EP/W02683X/1). We further acknowledge financial support from the UK funding agency EPSRC (grants EP/W007444/1, EP/V035975/1,  EP/V000624/1, EP/X009491/1 and EP/T028424/1), the Leverhulme Trust (RPG-2022-57 and RPG-2018-266), the Royal Society Wolfson Fellowship (RSWF/R3/183013), the Department for the Economy Northern Ireland under the US-Ireland R\&D Partnership Programme, the EU Horizon 2020 FET-Open project TeQ (766900),  the PNRR PE Italian National Quantum Science and Technology Institute (PE0000023) and the EU Horizon Europe EIC Pathfinder project QuCoM (GA no.10032223). We further acknowledge support from the QuantERA grant LEMAQUME, funded by the QuantERA II ERA-NET Cofund in Quantum Technologies implemented within the EU Horizon 2020 Programme. Data is available at: DOI: 10.5258/SOTON/D3337.

\bibliography{apssamp}

\providecommand{\noopsort}[1]{}\providecommand{\singleletter}[1]{#1}%
\begin{thebibliography}{52}%
\makeatletter
\providecommand \@ifxundefined [1]{%
 \@ifx{#1\undefined}
}%
\providecommand \@ifnum [1]{%
 \ifnum #1\expandafter \@firstoftwo
 \else \expandafter \@secondoftwo
 \fi
}%
\providecommand \@ifx [1]{%
 \ifx #1\expandafter \@firstoftwo
 \else \expandafter \@secondoftwo
 \fi
}%
\providecommand \natexlab [1]{#1}%
\providecommand \enquote  [1]{``#1''}%
\providecommand \bibnamefont  [1]{#1}%
\providecommand \bibfnamefont [1]{#1}%
\providecommand \citenamefont [1]{#1}%
\providecommand \href@noop [0]{\@secondoftwo}%
\providecommand \href [0]{\begingroup \@sanitize@url \@href}%
\providecommand \@href[1]{\@@startlink{#1}\@@href}%
\providecommand \@@href[1]{\endgroup#1\@@endlink}%
\providecommand \@sanitize@url [0]{\catcode `\\12\catcode `\$12\catcode `\&12\catcode `\#12\catcode `\^12\catcode `\_12\catcode `\%12\relax}%
\providecommand \@@startlink[1]{}%
\providecommand \@@endlink[0]{}%
\providecommand \url  [0]{\begingroup\@sanitize@url \@url }%
\providecommand \@url [1]{\endgroup\@href {#1}{\urlprefix }}%
\providecommand \urlprefix  [0]{URL }%
\providecommand \Eprint [0]{\href }%
\providecommand \doibase [0]{https://doi.org/}%
\providecommand \selectlanguage [0]{\@gobble}%
\providecommand \bibinfo  [0]{\@secondoftwo}%
\providecommand \bibfield  [0]{\@secondoftwo}%
\providecommand \translation [1]{[#1]}%
\providecommand \BibitemOpen [0]{}%
\providecommand \bibitemStop [0]{}%
\providecommand \bibitemNoStop [0]{.\EOS\space}%
\providecommand \EOS [0]{\spacefactor3000\relax}%
\providecommand \BibitemShut  [1]{\csname bibitem#1\endcsname}%
\let\auto@bib@innerbib\@empty
%</preamble>
\bibitem [{\citenamefont {Arndt}\ and\ \citenamefont {Hornberger}(2014)}]{arndt2014testing}%
  \BibitemOpen
  \bibfield  {author} {\bibinfo {author} {\bibfnamefont {M.}~\bibnamefont {Arndt}}\ and\ \bibinfo {author} {\bibfnamefont {K.}~\bibnamefont {Hornberger}},\ }\bibfield  {title} {\bibinfo {title} {Testing the limits of quantum mechanical superpositions},\ }\href@noop {} {\bibfield  {journal} {\bibinfo  {journal} {Nature Physics}\ }\textbf {\bibinfo {volume} {10}},\ \bibinfo {pages} {271} (\bibinfo {year} {2014})}\BibitemShut {NoStop}%
\bibitem [{\citenamefont {Carney}\ \emph {et~al.}(2019)\citenamefont {Carney}, \citenamefont {Stamp},\ and\ \citenamefont {Taylor}}]{carney2019tabletop}%
  \BibitemOpen
  \bibfield  {author} {\bibinfo {author} {\bibfnamefont {D.}~\bibnamefont {Carney}}, \bibinfo {author} {\bibfnamefont {P.~C.}\ \bibnamefont {Stamp}},\ and\ \bibinfo {author} {\bibfnamefont {J.~M.}\ \bibnamefont {Taylor}},\ }\bibfield  {title} {\bibinfo {title} {Tabletop experiments for quantum gravity: a user’s manual},\ }\href@noop {} {\bibfield  {journal} {\bibinfo  {journal} {Classical and Quantum Gravity}\ }\textbf {\bibinfo {volume} {36}},\ \bibinfo {pages} {034001} (\bibinfo {year} {2019})}\BibitemShut {NoStop}%
\bibitem [{\citenamefont {Nimmrichter}\ and\ \citenamefont {Hornberger}(2013)}]{nimmrichter2013macroscopicity}%
  \BibitemOpen
  \bibfield  {author} {\bibinfo {author} {\bibfnamefont {S.}~\bibnamefont {Nimmrichter}}\ and\ \bibinfo {author} {\bibfnamefont {K.}~\bibnamefont {Hornberger}},\ }\bibfield  {title} {\bibinfo {title} {Macroscopicity of mechanical quantum superposition states},\ }\href@noop {} {\bibfield  {journal} {\bibinfo  {journal} {Physical Review Letters}\ }\textbf {\bibinfo {volume} {110}},\ \bibinfo {pages} {160403} (\bibinfo {year} {2013})}\BibitemShut {NoStop}%
\bibitem [{\citenamefont {Gonzalez-Ballestero}\ \emph {et~al.}(2021)\citenamefont {Gonzalez-Ballestero}, \citenamefont {Aspelmeyer}, \citenamefont {Novotny}, \citenamefont {Quidant},\ and\ \citenamefont {Romero-Isart}}]{gonzalez2021levitodynamics}%
  \BibitemOpen
  \bibfield  {author} {\bibinfo {author} {\bibfnamefont {C.}~\bibnamefont {Gonzalez-Ballestero}}, \bibinfo {author} {\bibfnamefont {M.}~\bibnamefont {Aspelmeyer}}, \bibinfo {author} {\bibfnamefont {L.}~\bibnamefont {Novotny}}, \bibinfo {author} {\bibfnamefont {R.}~\bibnamefont {Quidant}},\ and\ \bibinfo {author} {\bibfnamefont {O.}~\bibnamefont {Romero-Isart}},\ }\bibfield  {title} {\bibinfo {title} {Levitodynamics: Levitation and control of microscopic objects in vacuum},\ }\href@noop {} {\bibfield  {journal} {\bibinfo  {journal} {Science}\ }\textbf {\bibinfo {volume} {374}},\ \bibinfo {pages} {eabg3027} (\bibinfo {year} {2021})}\BibitemShut {NoStop}%
\bibitem [{\citenamefont {Roda-Llordes}\ \emph {et~al.}(2023)\citenamefont {Roda-Llordes}, \citenamefont {Riera-Campeny}, \citenamefont {Candoli}, \citenamefont {Grochowski},\ and\ \citenamefont {Romero-Isart}}]{rodamacroscopic}%
  \BibitemOpen
  \bibfield  {author} {\bibinfo {author} {\bibfnamefont {M.}~\bibnamefont {Roda-Llordes}}, \bibinfo {author} {\bibfnamefont {A.}~\bibnamefont {Riera-Campeny}}, \bibinfo {author} {\bibfnamefont {D.}~\bibnamefont {Candoli}}, \bibinfo {author} {\bibfnamefont {P.~T.}\ \bibnamefont {Grochowski}},\ and\ \bibinfo {author} {\bibfnamefont {O.}~\bibnamefont {Romero-Isart}},\ }\bibfield  {title} {\bibinfo {title} {Macroscopic quantum superpositions via dynamics in a wide double-well potential},\ }\href@noop {} {\bibfield  {journal} {\bibinfo  {journal} {arXiv preprint arXiv:2303.07959}\ } (\bibinfo {year} {2023})}\BibitemShut {NoStop}%
\bibitem [{\citenamefont {Rakhubovsky}\ \emph {et~al.}(2019)\citenamefont {Rakhubovsky}, \citenamefont {Moore},\ and\ \citenamefont {Filip}}]{rakhubovsky2019nonclassical}%
  \BibitemOpen
  \bibfield  {author} {\bibinfo {author} {\bibfnamefont {A.~A.}\ \bibnamefont {Rakhubovsky}}, \bibinfo {author} {\bibfnamefont {D.~W.}\ \bibnamefont {Moore}},\ and\ \bibinfo {author} {\bibfnamefont {R.}~\bibnamefont {Filip}},\ }\bibfield  {title} {\bibinfo {title} {Nonclassical states of levitated macroscopic objects beyond the ground state},\ }\href@noop {} {\bibfield  {journal} {\bibinfo  {journal} {Quantum Science and Technology}\ }\textbf {\bibinfo {volume} {4}},\ \bibinfo {pages} {024006} (\bibinfo {year} {2019})}\BibitemShut {NoStop}%
\bibitem [{\citenamefont {Bonvin}\ \emph {et~al.}(2023)\citenamefont {Bonvin}, \citenamefont {Devaud}, \citenamefont {Rossi}, \citenamefont {Militaru}, \citenamefont {Dania}, \citenamefont {Bykov}, \citenamefont {Romero-Isart}, \citenamefont {Northup}, \citenamefont {Novotny},\ and\ \citenamefont {Frimmer}}]{bonvin2023state}%
  \BibitemOpen
  \bibfield  {author} {\bibinfo {author} {\bibfnamefont {E.}~\bibnamefont {Bonvin}}, \bibinfo {author} {\bibfnamefont {L.}~\bibnamefont {Devaud}}, \bibinfo {author} {\bibfnamefont {M.}~\bibnamefont {Rossi}}, \bibinfo {author} {\bibfnamefont {A.}~\bibnamefont {Militaru}}, \bibinfo {author} {\bibfnamefont {L.}~\bibnamefont {Dania}}, \bibinfo {author} {\bibfnamefont {D.~S.}\ \bibnamefont {Bykov}}, \bibinfo {author} {\bibfnamefont {O.}~\bibnamefont {Romero-Isart}}, \bibinfo {author} {\bibfnamefont {T.~E.}\ \bibnamefont {Northup}}, \bibinfo {author} {\bibfnamefont {L.}~\bibnamefont {Novotny}},\ and\ \bibinfo {author} {\bibfnamefont {M.}~\bibnamefont {Frimmer}},\ }\bibfield  {title} {\bibinfo {title} {State expansion of a levitated nanoparticle in a dark harmonic potential},\ }\href@noop {} {\bibfield  {journal} {\bibinfo  {journal} {arXiv:2312.13111}\ } (\bibinfo {year} {2023})}\BibitemShut {NoStop}%
\bibitem [{\citenamefont {Rashid}\ \emph {et~al.}(2016)\citenamefont {Rashid}, \citenamefont {Tufarelli}, \citenamefont {Bateman}, \citenamefont {Vovrosh}, \citenamefont {Hempston}, \citenamefont {Kim},\ and\ \citenamefont {Ulbricht}}]{rashid2016experimental}%
  \BibitemOpen
  \bibfield  {author} {\bibinfo {author} {\bibfnamefont {M.}~\bibnamefont {Rashid}}, \bibinfo {author} {\bibfnamefont {T.}~\bibnamefont {Tufarelli}}, \bibinfo {author} {\bibfnamefont {J.}~\bibnamefont {Bateman}}, \bibinfo {author} {\bibfnamefont {J.}~\bibnamefont {Vovrosh}}, \bibinfo {author} {\bibfnamefont {D.}~\bibnamefont {Hempston}}, \bibinfo {author} {\bibfnamefont {M.}~\bibnamefont {Kim}},\ and\ \bibinfo {author} {\bibfnamefont {H.}~\bibnamefont {Ulbricht}},\ }\bibfield  {title} {\bibinfo {title} {Experimental realization of a thermal squeezed state of levitated optomechanics},\ }\href@noop {} {\bibfield  {journal} {\bibinfo  {journal} {Physical Review Letters}\ }\textbf {\bibinfo {volume} {117}},\ \bibinfo {pages} {273601} (\bibinfo {year} {2016})}\BibitemShut {NoStop}%
\bibitem [{\citenamefont {Janszky}\ and\ \citenamefont {Yushin}(1986)}]{janszky1986squeezing}%
  \BibitemOpen
  \bibfield  {author} {\bibinfo {author} {\bibfnamefont {J.}~\bibnamefont {Janszky}}\ and\ \bibinfo {author} {\bibfnamefont {Y.~Y.}\ \bibnamefont {Yushin}},\ }\bibfield  {title} {\bibinfo {title} {Squeezing via frequency jump},\ }\href@noop {} {\bibfield  {journal} {\bibinfo  {journal} {Optics communications}\ }\textbf {\bibinfo {volume} {59}},\ \bibinfo {pages} {151} (\bibinfo {year} {1986})}\BibitemShut {NoStop}%
\bibitem [{\citenamefont {Penny}\ \emph {et~al.}(2023)\citenamefont {Penny}, \citenamefont {Pontin},\ and\ \citenamefont {Barker}}]{penny2023sympathetic}%
  \BibitemOpen
  \bibfield  {author} {\bibinfo {author} {\bibfnamefont {T.}~\bibnamefont {Penny}}, \bibinfo {author} {\bibfnamefont {A.}~\bibnamefont {Pontin}},\ and\ \bibinfo {author} {\bibfnamefont {P.}~\bibnamefont {Barker}},\ }\bibfield  {title} {\bibinfo {title} {Sympathetic cooling and squeezing of two colevitated nanoparticles},\ }\href@noop {} {\bibfield  {journal} {\bibinfo  {journal} {Physical Review Research}\ }\textbf {\bibinfo {volume} {5}},\ \bibinfo {pages} {013070} (\bibinfo {year} {2023})}\BibitemShut {NoStop}%
\bibitem [{\citenamefont {Li}\ \emph {et~al.}(2023)\citenamefont {Li}, \citenamefont {Xu}, \citenamefont {Huang},\ and\ \citenamefont {Liu}}]{li2023mechanical}%
  \BibitemOpen
  \bibfield  {author} {\bibinfo {author} {\bibfnamefont {Y.}~\bibnamefont {Li}}, \bibinfo {author} {\bibfnamefont {A.-N.}\ \bibnamefont {Xu}}, \bibinfo {author} {\bibfnamefont {L.-G.}\ \bibnamefont {Huang}},\ and\ \bibinfo {author} {\bibfnamefont {Y.-C.}\ \bibnamefont {Liu}},\ }\bibfield  {title} {\bibinfo {title} {Mechanical squeezing via detuning-switched driving},\ }\href@noop {} {\bibfield  {journal} {\bibinfo  {journal} {Physical Review A}\ }\textbf {\bibinfo {volume} {107}},\ \bibinfo {pages} {033508} (\bibinfo {year} {2023})}\BibitemShut {NoStop}%
\bibitem [{\citenamefont {Arnol'd}(2013)}]{arnol2013mathematical}%
  \BibitemOpen
  \bibfield  {author} {\bibinfo {author} {\bibfnamefont {V.~I.}\ \bibnamefont {Arnol'd}},\ }\href@noop {} {\emph {\bibinfo {title} {Mathematical methods of classical mechanics}}},\ Vol.~\bibinfo {volume} {60}\ (\bibinfo  {publisher} {Springer Science \& Business Media},\ \bibinfo {year} {2013})\BibitemShut {NoStop}%
\bibitem [{\citenamefont {Setter}\ \emph {et~al.}(2019)\citenamefont {Setter}, \citenamefont {Vovrosh},\ and\ \citenamefont {Ulbricht}}]{setter2019characterization}%
  \BibitemOpen
  \bibfield  {author} {\bibinfo {author} {\bibfnamefont {A.}~\bibnamefont {Setter}}, \bibinfo {author} {\bibfnamefont {J.}~\bibnamefont {Vovrosh}},\ and\ \bibinfo {author} {\bibfnamefont {H.}~\bibnamefont {Ulbricht}},\ }\bibfield  {title} {\bibinfo {title} {Characterization of non-linearities through mechanical squeezing in levitated optomechanics},\ }\href@noop {} {\bibfield  {journal} {\bibinfo  {journal} {Applied Physics Letters}\ }\textbf {\bibinfo {volume} {115}} (\bibinfo {year} {2019})}\BibitemShut {NoStop}%
\bibitem [{\citenamefont {Frimmer}\ \emph {et~al.}(2019)\citenamefont {Frimmer}, \citenamefont {Heugel}, \citenamefont {Nosan}, \citenamefont {Tebbenjohanns}, \citenamefont {H{\"a}lg}, \citenamefont {Akin}, \citenamefont {Degen}, \citenamefont {Novotny}, \citenamefont {Chitra}, \citenamefont {Zilberberg} \emph {et~al.}}]{frimmer2019rapid}%
  \BibitemOpen
  \bibfield  {author} {\bibinfo {author} {\bibfnamefont {M.}~\bibnamefont {Frimmer}}, \bibinfo {author} {\bibfnamefont {T.~L.}\ \bibnamefont {Heugel}}, \bibinfo {author} {\bibfnamefont {{\v{Z}}.}~\bibnamefont {Nosan}}, \bibinfo {author} {\bibfnamefont {F.}~\bibnamefont {Tebbenjohanns}}, \bibinfo {author} {\bibfnamefont {D.}~\bibnamefont {H{\"a}lg}}, \bibinfo {author} {\bibfnamefont {A.}~\bibnamefont {Akin}}, \bibinfo {author} {\bibfnamefont {C.~L.}\ \bibnamefont {Degen}}, \bibinfo {author} {\bibfnamefont {L.}~\bibnamefont {Novotny}}, \bibinfo {author} {\bibfnamefont {R.}~\bibnamefont {Chitra}}, \bibinfo {author} {\bibfnamefont {O.}~\bibnamefont {Zilberberg}}, \emph {et~al.},\ }\bibfield  {title} {\bibinfo {title} {Rapid flipping of parametric phase states},\ }\href@noop {} {\bibfield  {journal} {\bibinfo  {journal} {Physical review letters}\ }\textbf {\bibinfo {volume} {123}},\ \bibinfo {pages} {254102} (\bibinfo {year} {2019})}\BibitemShut {NoStop}%
\bibitem [{\citenamefont {Ricci}\ \emph {et~al.}(2017)\citenamefont {Ricci}, \citenamefont {Rica}, \citenamefont {Spasenovi{\'c}}, \citenamefont {Gieseler}, \citenamefont {Rondin}, \citenamefont {Novotny},\ and\ \citenamefont {Quidant}}]{ricci2017optically}%
  \BibitemOpen
  \bibfield  {author} {\bibinfo {author} {\bibfnamefont {F.}~\bibnamefont {Ricci}}, \bibinfo {author} {\bibfnamefont {R.~A.}\ \bibnamefont {Rica}}, \bibinfo {author} {\bibfnamefont {M.}~\bibnamefont {Spasenovi{\'c}}}, \bibinfo {author} {\bibfnamefont {J.}~\bibnamefont {Gieseler}}, \bibinfo {author} {\bibfnamefont {L.}~\bibnamefont {Rondin}}, \bibinfo {author} {\bibfnamefont {L.}~\bibnamefont {Novotny}},\ and\ \bibinfo {author} {\bibfnamefont {R.}~\bibnamefont {Quidant}},\ }\bibfield  {title} {\bibinfo {title} {Optically levitated nanoparticle as a model system for stochastic bistable dynamics},\ }\href@noop {} {\bibfield  {journal} {\bibinfo  {journal} {Nature communications}\ }\textbf {\bibinfo {volume} {8}},\ \bibinfo {pages} {15141} (\bibinfo {year} {2017})}\BibitemShut {NoStop}%
\bibitem [{\citenamefont {Gieseler}\ \emph {et~al.}(2014{\natexlab{a}})\citenamefont {Gieseler}, \citenamefont {Quidant}, \citenamefont {Dellago},\ and\ \citenamefont {Novotny}}]{gieseler2014dynamic}%
  \BibitemOpen
  \bibfield  {author} {\bibinfo {author} {\bibfnamefont {J.}~\bibnamefont {Gieseler}}, \bibinfo {author} {\bibfnamefont {R.}~\bibnamefont {Quidant}}, \bibinfo {author} {\bibfnamefont {C.}~\bibnamefont {Dellago}},\ and\ \bibinfo {author} {\bibfnamefont {L.}~\bibnamefont {Novotny}},\ }\bibfield  {title} {\bibinfo {title} {Dynamic relaxation of a levitated nanoparticle from a non-equilibrium steady state},\ }\href@noop {} {\bibfield  {journal} {\bibinfo  {journal} {Nature nanotechnology}\ }\textbf {\bibinfo {volume} {9}},\ \bibinfo {pages} {358} (\bibinfo {year} {2014}{\natexlab{a}})}\BibitemShut {NoStop}%
\bibitem [{\citenamefont {Gieseler}\ \emph {et~al.}(2014{\natexlab{b}})\citenamefont {Gieseler}, \citenamefont {Spasenovi{\'c}}, \citenamefont {Novotny},\ and\ \citenamefont {Quidant}}]{gieseler2014nonlinear}%
  \BibitemOpen
  \bibfield  {author} {\bibinfo {author} {\bibfnamefont {J.}~\bibnamefont {Gieseler}}, \bibinfo {author} {\bibfnamefont {M.}~\bibnamefont {Spasenovi{\'c}}}, \bibinfo {author} {\bibfnamefont {L.}~\bibnamefont {Novotny}},\ and\ \bibinfo {author} {\bibfnamefont {R.}~\bibnamefont {Quidant}},\ }\bibfield  {title} {\bibinfo {title} {Nonlinear mode coupling and synchronization of a vacuum-trapped nanoparticle},\ }\href@noop {} {\bibfield  {journal} {\bibinfo  {journal} {Physical review letters}\ }\textbf {\bibinfo {volume} {112}},\ \bibinfo {pages} {103603} (\bibinfo {year} {2014}{\natexlab{b}})}\BibitemShut {NoStop}%
\bibitem [{\citenamefont {Militaru}\ \emph {et~al.}(2021)\citenamefont {Militaru}, \citenamefont {Innerbichler}, \citenamefont {Frimmer}, \citenamefont {Tebbenjohanns}, \citenamefont {Novotny},\ and\ \citenamefont {Dellago}}]{militaru2021escape}%
  \BibitemOpen
  \bibfield  {author} {\bibinfo {author} {\bibfnamefont {A.}~\bibnamefont {Militaru}}, \bibinfo {author} {\bibfnamefont {M.}~\bibnamefont {Innerbichler}}, \bibinfo {author} {\bibfnamefont {M.}~\bibnamefont {Frimmer}}, \bibinfo {author} {\bibfnamefont {F.}~\bibnamefont {Tebbenjohanns}}, \bibinfo {author} {\bibfnamefont {L.}~\bibnamefont {Novotny}},\ and\ \bibinfo {author} {\bibfnamefont {C.}~\bibnamefont {Dellago}},\ }\bibfield  {title} {\bibinfo {title} {Escape dynamics of active particles in multistable potentials},\ }\href@noop {} {\bibfield  {journal} {\bibinfo  {journal} {Nature Communications}\ }\textbf {\bibinfo {volume} {12}},\ \bibinfo {pages} {2446} (\bibinfo {year} {2021})}\BibitemShut {NoStop}%
\bibitem [{\citenamefont {Cosco}\ \emph {et~al.}(2021)\citenamefont {Cosco}, \citenamefont {Pedernales},\ and\ \citenamefont {Plenio}}]{cosco2021enhanced}%
  \BibitemOpen
  \bibfield  {author} {\bibinfo {author} {\bibfnamefont {F.}~\bibnamefont {Cosco}}, \bibinfo {author} {\bibfnamefont {J.}~\bibnamefont {Pedernales}},\ and\ \bibinfo {author} {\bibfnamefont {M.~B.}\ \bibnamefont {Plenio}},\ }\bibfield  {title} {\bibinfo {title} {Enhanced force sensitivity and entanglement in periodically driven optomechanics},\ }\href@noop {} {\bibfield  {journal} {\bibinfo  {journal} {Physical Review A}\ }\textbf {\bibinfo {volume} {103}},\ \bibinfo {pages} {L061501} (\bibinfo {year} {2021})}\BibitemShut {NoStop}%
\bibitem [{\citenamefont {Tebbenjohanns}\ \emph {et~al.}(2021)\citenamefont {Tebbenjohanns}, \citenamefont {Mattana}, \citenamefont {Rossi}, \citenamefont {Frimmer},\ and\ \citenamefont {Novotny}}]{tebbenjohanns2021quantum}%
  \BibitemOpen
  \bibfield  {author} {\bibinfo {author} {\bibfnamefont {F.}~\bibnamefont {Tebbenjohanns}}, \bibinfo {author} {\bibfnamefont {M.~L.}\ \bibnamefont {Mattana}}, \bibinfo {author} {\bibfnamefont {M.}~\bibnamefont {Rossi}}, \bibinfo {author} {\bibfnamefont {M.}~\bibnamefont {Frimmer}},\ and\ \bibinfo {author} {\bibfnamefont {L.}~\bibnamefont {Novotny}},\ }\bibfield  {title} {\bibinfo {title} {Quantum control of a nanoparticle optically levitated in cryogenic free space},\ }\href@noop {} {\bibfield  {journal} {\bibinfo  {journal} {Nature}\ }\textbf {\bibinfo {volume} {595}},\ \bibinfo {pages} {378} (\bibinfo {year} {2021})}\BibitemShut {NoStop}%
\bibitem [{\citenamefont {Deli{\'c}}\ \emph {et~al.}(2020)\citenamefont {Deli{\'c}}, \citenamefont {Reisenbauer}, \citenamefont {Dare}, \citenamefont {Grass}, \citenamefont {Vuleti{\'c}}, \citenamefont {Kiesel},\ and\ \citenamefont {Aspelmeyer}}]{delic2020cooling}%
  \BibitemOpen
  \bibfield  {author} {\bibinfo {author} {\bibfnamefont {U.}~\bibnamefont {Deli{\'c}}}, \bibinfo {author} {\bibfnamefont {M.}~\bibnamefont {Reisenbauer}}, \bibinfo {author} {\bibfnamefont {K.}~\bibnamefont {Dare}}, \bibinfo {author} {\bibfnamefont {D.}~\bibnamefont {Grass}}, \bibinfo {author} {\bibfnamefont {V.}~\bibnamefont {Vuleti{\'c}}}, \bibinfo {author} {\bibfnamefont {N.}~\bibnamefont {Kiesel}},\ and\ \bibinfo {author} {\bibfnamefont {M.}~\bibnamefont {Aspelmeyer}},\ }\bibfield  {title} {\bibinfo {title} {Cooling of a levitated nanoparticle to the motional quantum ground state},\ }\href@noop {} {\bibfield  {journal} {\bibinfo  {journal} {Science}\ }\textbf {\bibinfo {volume} {367}},\ \bibinfo {pages} {892} (\bibinfo {year} {2020})}\BibitemShut {NoStop}%
\bibitem [{\citenamefont {Magrini}\ \emph {et~al.}(2021)\citenamefont {Magrini}, \citenamefont {Rosenzweig}, \citenamefont {Bach}, \citenamefont {Deutschmann-Olek}, \citenamefont {Hofer}, \citenamefont {Hong}, \citenamefont {Kiesel}, \citenamefont {Kugi},\ and\ \citenamefont {Aspelmeyer}}]{magrini2021real}%
  \BibitemOpen
  \bibfield  {author} {\bibinfo {author} {\bibfnamefont {L.}~\bibnamefont {Magrini}}, \bibinfo {author} {\bibfnamefont {P.}~\bibnamefont {Rosenzweig}}, \bibinfo {author} {\bibfnamefont {C.}~\bibnamefont {Bach}}, \bibinfo {author} {\bibfnamefont {A.}~\bibnamefont {Deutschmann-Olek}}, \bibinfo {author} {\bibfnamefont {S.~G.}\ \bibnamefont {Hofer}}, \bibinfo {author} {\bibfnamefont {S.}~\bibnamefont {Hong}}, \bibinfo {author} {\bibfnamefont {N.}~\bibnamefont {Kiesel}}, \bibinfo {author} {\bibfnamefont {A.}~\bibnamefont {Kugi}},\ and\ \bibinfo {author} {\bibfnamefont {M.}~\bibnamefont {Aspelmeyer}},\ }\bibfield  {title} {\bibinfo {title} {Real-time optimal quantum control of mechanical motion at room temperature},\ }\href@noop {} {\bibfield  {journal} {\bibinfo  {journal} {Nature}\ }\textbf {\bibinfo {volume} {595}},\ \bibinfo {pages} {373} (\bibinfo {year} {2021})}\BibitemShut {NoStop}%
\bibitem [{\citenamefont {Kamba}\ \emph {et~al.}(2022)\citenamefont {Kamba}, \citenamefont {Shimizu},\ and\ \citenamefont {Aikawa}}]{kamba2022optical}%
  \BibitemOpen
  \bibfield  {author} {\bibinfo {author} {\bibfnamefont {M.}~\bibnamefont {Kamba}}, \bibinfo {author} {\bibfnamefont {R.}~\bibnamefont {Shimizu}},\ and\ \bibinfo {author} {\bibfnamefont {K.}~\bibnamefont {Aikawa}},\ }\bibfield  {title} {\bibinfo {title} {Optical cold damping of neutral nanoparticles near the ground state in an optical lattice},\ }\href@noop {} {\bibfield  {journal} {\bibinfo  {journal} {Optics Express}\ }\textbf {\bibinfo {volume} {30}},\ \bibinfo {pages} {26716} (\bibinfo {year} {2022})}\BibitemShut {NoStop}%
\bibitem [{\citenamefont {Ranfagni}\ \emph {et~al.}(2022)\citenamefont {Ranfagni}, \citenamefont {B{\o}rkje}, \citenamefont {Marino},\ and\ \citenamefont {Marin}}]{ranfagni2022two}%
  \BibitemOpen
  \bibfield  {author} {\bibinfo {author} {\bibfnamefont {A.}~\bibnamefont {Ranfagni}}, \bibinfo {author} {\bibfnamefont {K.}~\bibnamefont {B{\o}rkje}}, \bibinfo {author} {\bibfnamefont {F.}~\bibnamefont {Marino}},\ and\ \bibinfo {author} {\bibfnamefont {F.}~\bibnamefont {Marin}},\ }\bibfield  {title} {\bibinfo {title} {Two-dimensional quantum motion of a levitated nanosphere},\ }\href@noop {} {\bibfield  {journal} {\bibinfo  {journal} {Physical Review Research}\ }\textbf {\bibinfo {volume} {4}},\ \bibinfo {pages} {033051} (\bibinfo {year} {2022})}\BibitemShut {NoStop}%
\bibitem [{\citenamefont {Gieseler}\ \emph {et~al.}(2013)\citenamefont {Gieseler}, \citenamefont {Novotny},\ and\ \citenamefont {Quidant}}]{gieseler2013thermal}%
  \BibitemOpen
  \bibfield  {author} {\bibinfo {author} {\bibfnamefont {J.}~\bibnamefont {Gieseler}}, \bibinfo {author} {\bibfnamefont {L.}~\bibnamefont {Novotny}},\ and\ \bibinfo {author} {\bibfnamefont {R.}~\bibnamefont {Quidant}},\ }\bibfield  {title} {\bibinfo {title} {Thermal nonlinearities in a nanomechanical oscillator},\ }\href@noop {} {\bibfield  {journal} {\bibinfo  {journal} {Nature Physics}\ }\textbf {\bibinfo {volume} {9}},\ \bibinfo {pages} {806} (\bibinfo {year} {2013})}\BibitemShut {NoStop}%
\bibitem [{\citenamefont {Jones}\ \emph {et~al.}(2015)\citenamefont {Jones}, \citenamefont {Marag{\`o}},\ and\ \citenamefont {Volpe}}]{jones2015optical}%
  \BibitemOpen
  \bibfield  {author} {\bibinfo {author} {\bibfnamefont {P.~H.}\ \bibnamefont {Jones}}, \bibinfo {author} {\bibfnamefont {O.~M.}\ \bibnamefont {Marag{\`o}}},\ and\ \bibinfo {author} {\bibfnamefont {G.}~\bibnamefont {Volpe}},\ }\href@noop {} {\emph {\bibinfo {title} {Optical tweezers: Principles and applications}}}\ (\bibinfo  {publisher} {Cambridge University Press},\ \bibinfo {year} {2015})\BibitemShut {NoStop}%
\bibitem [{\citenamefont {Ashkin}\ \emph {et~al.}(1986)\citenamefont {Ashkin}, \citenamefont {Dziedzic}, \citenamefont {Bjorkholm},\ and\ \citenamefont {Chu}}]{ashkin1986observation}%
  \BibitemOpen
  \bibfield  {author} {\bibinfo {author} {\bibfnamefont {A.}~\bibnamefont {Ashkin}}, \bibinfo {author} {\bibfnamefont {J.~M.}\ \bibnamefont {Dziedzic}}, \bibinfo {author} {\bibfnamefont {J.~E.}\ \bibnamefont {Bjorkholm}},\ and\ \bibinfo {author} {\bibfnamefont {S.}~\bibnamefont {Chu}},\ }\bibfield  {title} {\bibinfo {title} {Observation of a single-beam gradient force optical trap for dielectric particles},\ }\href@noop {} {\bibfield  {journal} {\bibinfo  {journal} {Optics Letters}\ }\textbf {\bibinfo {volume} {11}},\ \bibinfo {pages} {288} (\bibinfo {year} {1986})}\BibitemShut {NoStop}%
\bibitem [{\citenamefont {Harada}\ and\ \citenamefont {Asakura}(1996)}]{harada1996radiation}%
  \BibitemOpen
  \bibfield  {author} {\bibinfo {author} {\bibfnamefont {Y.}~\bibnamefont {Harada}}\ and\ \bibinfo {author} {\bibfnamefont {T.}~\bibnamefont {Asakura}},\ }\bibfield  {title} {\bibinfo {title} {Radiation forces on a dielectric sphere in the rayleigh scattering regime},\ }\href@noop {} {\bibfield  {journal} {\bibinfo  {journal} {Optics communications}\ }\textbf {\bibinfo {volume} {124}},\ \bibinfo {pages} {529} (\bibinfo {year} {1996})}\BibitemShut {NoStop}%
\bibitem [{\citenamefont {Kubo}(1966)}]{kubo1966fluctuation}%
  \BibitemOpen
  \bibfield  {author} {\bibinfo {author} {\bibfnamefont {R.}~\bibnamefont {Kubo}},\ }\bibfield  {title} {\bibinfo {title} {The fluctuation-dissipation theorem},\ }\href@noop {} {\bibfield  {journal} {\bibinfo  {journal} {Reports on Progress in Physics}\ }\textbf {\bibinfo {volume} {29}},\ \bibinfo {pages} {255} (\bibinfo {year} {1966})}\BibitemShut {NoStop}%
\bibitem [{\citenamefont {Volpe}\ and\ \citenamefont {Volpe}(2013)}]{volpe2013simulation}%
  \BibitemOpen
  \bibfield  {author} {\bibinfo {author} {\bibfnamefont {G.}~\bibnamefont {Volpe}}\ and\ \bibinfo {author} {\bibfnamefont {G.}~\bibnamefont {Volpe}},\ }\bibfield  {title} {\bibinfo {title} {Simulation of a brownian particle in an optical trap},\ }\href@noop {} {\bibfield  {journal} {\bibinfo  {journal} {American Journal of Physics}\ }\textbf {\bibinfo {volume} {81}},\ \bibinfo {pages} {224} (\bibinfo {year} {2013})}\BibitemShut {NoStop}%
\bibitem [{\citenamefont {Vovrosh}\ \emph {et~al.}(2017)\citenamefont {Vovrosh}, \citenamefont {Rashid}, \citenamefont {Hempston}, \citenamefont {Bateman}, \citenamefont {Paternostro},\ and\ \citenamefont {Ulbricht}}]{vovrosh2017parametric}%
  \BibitemOpen
  \bibfield  {author} {\bibinfo {author} {\bibfnamefont {J.}~\bibnamefont {Vovrosh}}, \bibinfo {author} {\bibfnamefont {M.}~\bibnamefont {Rashid}}, \bibinfo {author} {\bibfnamefont {D.}~\bibnamefont {Hempston}}, \bibinfo {author} {\bibfnamefont {J.}~\bibnamefont {Bateman}}, \bibinfo {author} {\bibfnamefont {M.}~\bibnamefont {Paternostro}},\ and\ \bibinfo {author} {\bibfnamefont {H.}~\bibnamefont {Ulbricht}},\ }\bibfield  {title} {\bibinfo {title} {Parametric feedback cooling of levitated optomechanics in a parabolic mirror trap},\ }\href@noop {} {\bibfield  {journal} {\bibinfo  {journal} {JOSA B}\ }\textbf {\bibinfo {volume} {34}},\ \bibinfo {pages} {1421} (\bibinfo {year} {2017})}\BibitemShut {NoStop}%
\bibitem [{\citenamefont {Dawson}\ and\ \citenamefont {Bateman}(2019)}]{dawson2019spectral}%
  \BibitemOpen
  \bibfield  {author} {\bibinfo {author} {\bibfnamefont {C.}~\bibnamefont {Dawson}}\ and\ \bibinfo {author} {\bibfnamefont {J.}~\bibnamefont {Bateman}},\ }\bibfield  {title} {\bibinfo {title} {Spectral analysis and parameter estimation in levitated optomechanics},\ }\href@noop {} {\bibfield  {journal} {\bibinfo  {journal} {JOSA B}\ }\textbf {\bibinfo {volume} {36}},\ \bibinfo {pages} {1565} (\bibinfo {year} {2019})}\BibitemShut {NoStop}%
\bibitem [{\citenamefont {Ashman}\ \emph {et~al.}(1994)\citenamefont {Ashman}, \citenamefont {Bird},\ and\ \citenamefont {Zepf}}]{ashman1994detecting}%
  \BibitemOpen
  \bibfield  {author} {\bibinfo {author} {\bibfnamefont {K.~M.}\ \bibnamefont {Ashman}}, \bibinfo {author} {\bibfnamefont {C.~M.}\ \bibnamefont {Bird}},\ and\ \bibinfo {author} {\bibfnamefont {S.~E.}\ \bibnamefont {Zepf}},\ }\bibfield  {title} {\bibinfo {title} {Detecting bimodality in astronomical datasets},\ }\href@noop {} {\bibfield  {journal} {\bibinfo  {journal} {Astron. J.}\ }\textbf {\bibinfo {volume} {108}},\ \bibinfo {pages} {2348} (\bibinfo {year} {1994})}\BibitemShut {NoStop}%
\bibitem [{\citenamefont {Bykov}\ \emph {et~al.}(2019)\citenamefont {Bykov}, \citenamefont {Mestres}, \citenamefont {Dania}, \citenamefont {Schmöger},\ and\ \citenamefont {Northup}}]{Bykov_2019}%
  \BibitemOpen
  \bibfield  {author} {\bibinfo {author} {\bibfnamefont {D.~S.}\ \bibnamefont {Bykov}}, \bibinfo {author} {\bibfnamefont {P.}~\bibnamefont {Mestres}}, \bibinfo {author} {\bibfnamefont {L.}~\bibnamefont {Dania}}, \bibinfo {author} {\bibfnamefont {L.}~\bibnamefont {Schmöger}},\ and\ \bibinfo {author} {\bibfnamefont {T.~E.}\ \bibnamefont {Northup}},\ }\bibfield  {title} {\bibinfo {title} {{Direct loading of nanoparticles under high vacuum into a Paul trap for levitodynamical experiments}},\ }\href@noop {} {\bibfield  {journal} {\bibinfo  {journal} {Applied Physics Letters}\ }\textbf {\bibinfo {volume} {115}},\ \bibinfo {pages} {034101} (\bibinfo {year} {2019})}\BibitemShut {NoStop}%
\bibitem [{\citenamefont {Weiss}\ \emph {et~al.}(2021)\citenamefont {Weiss}, \citenamefont {Roda-Llordes}, \citenamefont {Torrontegui}, \citenamefont {Aspelmeyer},\ and\ \citenamefont {Romero-Isart}}]{Weiss_2021}%
  \BibitemOpen
  \bibfield  {author} {\bibinfo {author} {\bibfnamefont {T.}~\bibnamefont {Weiss}}, \bibinfo {author} {\bibfnamefont {M.}~\bibnamefont {Roda-Llordes}}, \bibinfo {author} {\bibfnamefont {E.}~\bibnamefont {Torrontegui}}, \bibinfo {author} {\bibfnamefont {M.}~\bibnamefont {Aspelmeyer}},\ and\ \bibinfo {author} {\bibfnamefont {O.}~\bibnamefont {Romero-Isart}},\ }\bibfield  {title} {\bibinfo {title} {Large quantum delocalization of a levitated nanoparticle using optimal control: Applications for force sensing and entangling via weak forces},\ }\href@noop {} {\bibfield  {journal} {\bibinfo  {journal} {Phys. Rev. Lett.}\ }\textbf {\bibinfo {volume} {127}},\ \bibinfo {pages} {023601} (\bibinfo {year} {2021})}\BibitemShut {NoStop}%
\bibitem [{\citenamefont {Gonzalez-Ballestero}\ \emph {et~al.}(2019)\citenamefont {Gonzalez-Ballestero}, \citenamefont {Maurer}, \citenamefont {Windey}, \citenamefont {Novotny}, \citenamefont {Reimann},\ and\ \citenamefont {Romero-Isart}}]{Gonzalez-Ballestero_2019}%
  \BibitemOpen
  \bibfield  {author} {\bibinfo {author} {\bibfnamefont {C.}~\bibnamefont {Gonzalez-Ballestero}}, \bibinfo {author} {\bibfnamefont {P.}~\bibnamefont {Maurer}}, \bibinfo {author} {\bibfnamefont {D.}~\bibnamefont {Windey}}, \bibinfo {author} {\bibfnamefont {L.}~\bibnamefont {Novotny}}, \bibinfo {author} {\bibfnamefont {R.}~\bibnamefont {Reimann}},\ and\ \bibinfo {author} {\bibfnamefont {O.}~\bibnamefont {Romero-Isart}},\ }\bibfield  {title} {\bibinfo {title} {Theory for cavity cooling of levitated nanoparticles via coherent scattering: Master equation approach},\ }\href@noop {} {\bibfield  {journal} {\bibinfo  {journal} {Phys. Rev. A}\ }\textbf {\bibinfo {volume} {100}},\ \bibinfo {pages} {013805} (\bibinfo {year} {2019})}\BibitemShut {NoStop}%
\bibitem [{\citenamefont {Seberson}\ and\ \citenamefont {Robicheaux}(2020)}]{Seberson_2020}%
  \BibitemOpen
  \bibfield  {author} {\bibinfo {author} {\bibfnamefont {T.}~\bibnamefont {Seberson}}\ and\ \bibinfo {author} {\bibfnamefont {F.}~\bibnamefont {Robicheaux}},\ }\bibfield  {title} {\bibinfo {title} {Distribution of laser shot-noise energy delivered to a levitated nanoparticle},\ }\href@noop {} {\bibfield  {journal} {\bibinfo  {journal} {Phys. Rev. A}\ }\textbf {\bibinfo {volume} {102}},\ \bibinfo {pages} {033505} (\bibinfo {year} {2020})}\BibitemShut {NoStop}%
\bibitem [{Note1()}]{Note1}%
  \BibitemOpen
  \bibinfo {note} {We take the particle radius $R=4$\protect \,nm, the laser wavelength $\lambda =1550$\protect \,nm, the laser power $P_0=0.5$\protect \,W, the trap width $w_0=750$\protect \,nm, the vacuum permittivity $\varepsilon _0=8.9\times 10^{-21}$\protect \,F/nm, the relative dielectric constant $\epsilon =2$ and the asymmetry factors $A_x=1$, $A_y=0.9$.}\BibitemShut {Stop}%
\bibitem [{Note2()}]{Note2}%
  \BibitemOpen
  \bibinfo {note} {We take mean excitation number $\mathinner {\langle {n}\rangle }=(\exp [\hbar \omega /k_BT]-1)^{-1}=4.52$. Here $\omega $ is the oscillation frequency approximated at the low-energy section of our inverted Gaussian system (\protect \textit {e.g.}~in our example $\omega /2\pi =80$\protect \,kHz). The variance of this thermal state is half of that of the blurred Fock state, as shown by Fig.~\ref {fig:quantum_simulation}H.}\BibitemShut {Stop}%
\bibitem [{\citenamefont {Hempston}\ \emph {et~al.}(2017)\citenamefont {Hempston}, \citenamefont {Vovrosh}, \citenamefont {Toroš}, \citenamefont {Winstone}, \citenamefont {Rashid},\ and\ \citenamefont {Ulbricht}}]{Hempston_2017}%
  \BibitemOpen
  \bibfield  {author} {\bibinfo {author} {\bibfnamefont {D.}~\bibnamefont {Hempston}}, \bibinfo {author} {\bibfnamefont {J.}~\bibnamefont {Vovrosh}}, \bibinfo {author} {\bibfnamefont {M.}~\bibnamefont {Toroš}}, \bibinfo {author} {\bibfnamefont {G.}~\bibnamefont {Winstone}}, \bibinfo {author} {\bibfnamefont {M.}~\bibnamefont {Rashid}},\ and\ \bibinfo {author} {\bibfnamefont {H.}~\bibnamefont {Ulbricht}},\ }\bibfield  {title} {\bibinfo {title} {{Force sensing with an optically levitated charged nanoparticle}},\ }\href@noop {} {\bibfield  {journal} {\bibinfo  {journal} {Applied Physics Letters}\ }\textbf {\bibinfo {volume} {111}},\ \bibinfo {pages} {133111} (\bibinfo {year} {2017})}\BibitemShut {NoStop}%
\bibitem [{\citenamefont {Kremer}\ \emph {et~al.}(2024)\citenamefont {Kremer}, \citenamefont {Tandeitnik}, \citenamefont {Mufato}, \citenamefont {Califrer}, \citenamefont {Calderoni}, \citenamefont {Calliari}, \citenamefont {Melo}, \citenamefont {Tempor\~ao},\ and\ \citenamefont {Guerreiro}}]{Kremer_2024}%
  \BibitemOpen
  \bibfield  {author} {\bibinfo {author} {\bibfnamefont {O.}~\bibnamefont {Kremer}}, \bibinfo {author} {\bibfnamefont {D.}~\bibnamefont {Tandeitnik}}, \bibinfo {author} {\bibfnamefont {R.}~\bibnamefont {Mufato}}, \bibinfo {author} {\bibfnamefont {I.}~\bibnamefont {Califrer}}, \bibinfo {author} {\bibfnamefont {B.}~\bibnamefont {Calderoni}}, \bibinfo {author} {\bibfnamefont {F.}~\bibnamefont {Calliari}}, \bibinfo {author} {\bibfnamefont {B.}~\bibnamefont {Melo}}, \bibinfo {author} {\bibfnamefont {G.}~\bibnamefont {Tempor\~ao}},\ and\ \bibinfo {author} {\bibfnamefont {T.}~\bibnamefont {Guerreiro}},\ }\bibfield  {title} {\bibinfo {title} {Perturbative nonlinear feedback forces for optical levitation experiments},\ }\href@noop {} {\bibfield  {journal} {\bibinfo  {journal} {Phys. Rev. A}\ }\textbf {\bibinfo {volume} {109}},\ \bibinfo {pages} {023521} (\bibinfo {year} {2024})}\BibitemShut {NoStop}%
\bibitem [{\citenamefont {Szigeti}\ \emph {et~al.}(2014)\citenamefont {Szigeti}, \citenamefont {Carvalho}, \citenamefont {Morley},\ and\ \citenamefont {Hush}}]{Szigeti_2014}%
  \BibitemOpen
  \bibfield  {author} {\bibinfo {author} {\bibfnamefont {S.~S.}\ \bibnamefont {Szigeti}}, \bibinfo {author} {\bibfnamefont {A.~R.~R.}\ \bibnamefont {Carvalho}}, \bibinfo {author} {\bibfnamefont {J.~G.}\ \bibnamefont {Morley}},\ and\ \bibinfo {author} {\bibfnamefont {M.~R.}\ \bibnamefont {Hush}},\ }\bibfield  {title} {\bibinfo {title} {Ignorance is bliss: General and robust cancellation of decoherence via no-knowledge quantum feedback},\ }\href@noop {} {\bibfield  {journal} {\bibinfo  {journal} {Phys. Rev. Lett.}\ }\textbf {\bibinfo {volume} {113}},\ \bibinfo {pages} {020407} (\bibinfo {year} {2014})}\BibitemShut {NoStop}%
\bibitem [{\citenamefont {Gajewski}\ and\ \citenamefont {Bateman}(2024)}]{gajewski2024backaction}%
  \BibitemOpen
  \bibfield  {author} {\bibinfo {author} {\bibfnamefont {R.}~\bibnamefont {Gajewski}}\ and\ \bibinfo {author} {\bibfnamefont {J.}~\bibnamefont {Bateman}},\ }\bibfield  {title} {\bibinfo {title} {Backaction suppression in levitated optomechanics using reflective boundaries},\ }\href@noop {} {\bibfield  {journal} {\bibinfo  {journal} {arXiv preprint arXiv:2405.04366}\ } (\bibinfo {year} {2024})}\BibitemShut {NoStop}%
\bibitem [{\citenamefont {Weiser}\ \emph {et~al.}(2024)\citenamefont {Weiser}, \citenamefont {Faorlin}, \citenamefont {Panzl}, \citenamefont {Lafenthaler}, \citenamefont {Dania}, \citenamefont {Bykov}, \citenamefont {Monz}, \citenamefont {Blatt},\ and\ \citenamefont {Cerchiari}}]{weiser2024back}%
  \BibitemOpen
  \bibfield  {author} {\bibinfo {author} {\bibfnamefont {Y.}~\bibnamefont {Weiser}}, \bibinfo {author} {\bibfnamefont {T.}~\bibnamefont {Faorlin}}, \bibinfo {author} {\bibfnamefont {L.}~\bibnamefont {Panzl}}, \bibinfo {author} {\bibfnamefont {T.}~\bibnamefont {Lafenthaler}}, \bibinfo {author} {\bibfnamefont {L.}~\bibnamefont {Dania}}, \bibinfo {author} {\bibfnamefont {D.~S.}\ \bibnamefont {Bykov}}, \bibinfo {author} {\bibfnamefont {T.}~\bibnamefont {Monz}}, \bibinfo {author} {\bibfnamefont {R.}~\bibnamefont {Blatt}},\ and\ \bibinfo {author} {\bibfnamefont {G.}~\bibnamefont {Cerchiari}},\ }\bibfield  {title} {\bibinfo {title} {Back action suppression for levitated dipolar scatterers},\ }\href@noop {} {\bibfield  {journal} {\bibinfo  {journal} {arXiv preprint arXiv:2402.04802}\ } (\bibinfo {year} {2024})}\BibitemShut {NoStop}%
\bibitem [{\citenamefont {Vanner}\ \emph {et~al.}(2013)\citenamefont {Vanner}, \citenamefont {Hofer}, \citenamefont {Cole},\ and\ \citenamefont {Aspelmeyer}}]{vanner2013cooling}%
  \BibitemOpen
  \bibfield  {author} {\bibinfo {author} {\bibfnamefont {M.}~\bibnamefont {Vanner}}, \bibinfo {author} {\bibfnamefont {J.}~\bibnamefont {Hofer}}, \bibinfo {author} {\bibfnamefont {G.}~\bibnamefont {Cole}},\ and\ \bibinfo {author} {\bibfnamefont {M.}~\bibnamefont {Aspelmeyer}},\ }\bibfield  {title} {\bibinfo {title} {Cooling-by-measurement and mechanical state tomography via pulsed optomechanics},\ }\href@noop {} {\bibfield  {journal} {\bibinfo  {journal} {Nature communications}\ }\textbf {\bibinfo {volume} {4}},\ \bibinfo {pages} {2295} (\bibinfo {year} {2013})}\BibitemShut {NoStop}%
\bibitem [{\citenamefont {Rashid}\ \emph {et~al.}(2017)\citenamefont {Rashid}, \citenamefont {Toro{\v{s}}},\ and\ \citenamefont {Ulbricht}}]{rashid2017wigner}%
  \BibitemOpen
  \bibfield  {author} {\bibinfo {author} {\bibfnamefont {M.}~\bibnamefont {Rashid}}, \bibinfo {author} {\bibfnamefont {M.}~\bibnamefont {Toro{\v{s}}}},\ and\ \bibinfo {author} {\bibfnamefont {H.}~\bibnamefont {Ulbricht}},\ }\bibfield  {title} {\bibinfo {title} {Wigner function reconstruction in levitated optomechanics},\ }\href@noop {} {\bibfield  {journal} {\bibinfo  {journal} {Quantum Measurements and Quantum Metrology}\ }\textbf {\bibinfo {volume} {4}},\ \bibinfo {pages} {17} (\bibinfo {year} {2017})}\BibitemShut {NoStop}%
\bibitem [{\citenamefont {Vanner}\ \emph {et~al.}(2015)\citenamefont {Vanner}, \citenamefont {Pikovski},\ and\ \citenamefont {Kim}}]{vanner2015towards}%
  \BibitemOpen
  \bibfield  {author} {\bibinfo {author} {\bibfnamefont {M.~R.}\ \bibnamefont {Vanner}}, \bibinfo {author} {\bibfnamefont {I.}~\bibnamefont {Pikovski}},\ and\ \bibinfo {author} {\bibfnamefont {M.}~\bibnamefont {Kim}},\ }\bibfield  {title} {\bibinfo {title} {Towards optomechanical quantum state reconstruction of mechanical motion},\ }\href@noop {} {\bibfield  {journal} {\bibinfo  {journal} {Annalen der Physik}\ }\textbf {\bibinfo {volume} {527}},\ \bibinfo {pages} {15} (\bibinfo {year} {2015})}\BibitemShut {NoStop}%
\bibitem [{\citenamefont {Shahandeh}\ and\ \citenamefont {Ringbauer}(2019)}]{shahandeh2019optomechanical}%
  \BibitemOpen
  \bibfield  {author} {\bibinfo {author} {\bibfnamefont {F.}~\bibnamefont {Shahandeh}}\ and\ \bibinfo {author} {\bibfnamefont {M.}~\bibnamefont {Ringbauer}},\ }\bibfield  {title} {\bibinfo {title} {Optomechanical state reconstruction and nonclassicality verification beyond the resolved-sideband regime},\ }\href@noop {} {\bibfield  {journal} {\bibinfo  {journal} {Quantum}\ }\textbf {\bibinfo {volume} {3}},\ \bibinfo {pages} {125} (\bibinfo {year} {2019})}\BibitemShut {NoStop}%
\bibitem [{\citenamefont {Warszawski}\ \emph {et~al.}(2019)\citenamefont {Warszawski}, \citenamefont {Szorkovszky}, \citenamefont {Bowen},\ and\ \citenamefont {Doherty}}]{warszawski2019tomography}%
  \BibitemOpen
  \bibfield  {author} {\bibinfo {author} {\bibfnamefont {P.}~\bibnamefont {Warszawski}}, \bibinfo {author} {\bibfnamefont {A.}~\bibnamefont {Szorkovszky}}, \bibinfo {author} {\bibfnamefont {W.~P.}\ \bibnamefont {Bowen}},\ and\ \bibinfo {author} {\bibfnamefont {A.~C.}\ \bibnamefont {Doherty}},\ }\bibfield  {title} {\bibinfo {title} {Tomography of an optomechanical oscillator via parametrically amplified position measurement},\ }\href@noop {} {\bibfield  {journal} {\bibinfo  {journal} {New Journal of Physics}\ }\textbf {\bibinfo {volume} {21}},\ \bibinfo {pages} {023020} (\bibinfo {year} {2019})}\BibitemShut {NoStop}%
\bibitem [{\citenamefont {Weiss}\ and\ \citenamefont {Romero-Isart}(2019)}]{weiss2019quantum}%
  \BibitemOpen
  \bibfield  {author} {\bibinfo {author} {\bibfnamefont {T.}~\bibnamefont {Weiss}}\ and\ \bibinfo {author} {\bibfnamefont {O.}~\bibnamefont {Romero-Isart}},\ }\bibfield  {title} {\bibinfo {title} {Quantum motional state tomography with nonquadratic potentials and neural networks},\ }\href@noop {} {\bibfield  {journal} {\bibinfo  {journal} {Physical Review Research}\ }\textbf {\bibinfo {volume} {1}},\ \bibinfo {pages} {033157} (\bibinfo {year} {2019})}\BibitemShut {NoStop}%
\bibitem [{\citenamefont {Wu}\ \emph {et~al.}(2023)\citenamefont {Wu}, \citenamefont {Ciampini}, \citenamefont {Paternostro},\ and\ \citenamefont {Carlesso}}]{Qiongyuan_2023}%
  \BibitemOpen
  \bibfield  {author} {\bibinfo {author} {\bibfnamefont {Q.}~\bibnamefont {Wu}}, \bibinfo {author} {\bibfnamefont {M.~A.}\ \bibnamefont {Ciampini}}, \bibinfo {author} {\bibfnamefont {M.}~\bibnamefont {Paternostro}},\ and\ \bibinfo {author} {\bibfnamefont {M.}~\bibnamefont {Carlesso}},\ }\bibfield  {title} {\bibinfo {title} {Quantifying protocol efficiency: A thermodynamic figure of merit for classical and quantum state-transfer protocols},\ }\href@noop {} {\bibfield  {journal} {\bibinfo  {journal} {Phys. Rev. Res.}\ }\textbf {\bibinfo {volume} {5}},\ \bibinfo {pages} {023117} (\bibinfo {year} {2023})}\BibitemShut {NoStop}%
\bibitem [{\citenamefont {Marti}\ \emph {et~al.}(2024)\citenamefont {Marti}, \citenamefont {von L{\"u}pke}, \citenamefont {Joshi}, \citenamefont {Yang}, \citenamefont {Bild}, \citenamefont {Omahen}, \citenamefont {Chu},\ and\ \citenamefont {Fadel}}]{marti2024quantum}%
  \BibitemOpen
  \bibfield  {author} {\bibinfo {author} {\bibfnamefont {S.}~\bibnamefont {Marti}}, \bibinfo {author} {\bibfnamefont {U.}~\bibnamefont {von L{\"u}pke}}, \bibinfo {author} {\bibfnamefont {O.}~\bibnamefont {Joshi}}, \bibinfo {author} {\bibfnamefont {Y.}~\bibnamefont {Yang}}, \bibinfo {author} {\bibfnamefont {M.}~\bibnamefont {Bild}}, \bibinfo {author} {\bibfnamefont {A.}~\bibnamefont {Omahen}}, \bibinfo {author} {\bibfnamefont {Y.}~\bibnamefont {Chu}},\ and\ \bibinfo {author} {\bibfnamefont {M.}~\bibnamefont {Fadel}},\ }\bibfield  {title} {\bibinfo {title} {Quantum squeezing in a nonlinear mechanical oscillator},\ }\href@noop {} {\bibfield  {journal} {\bibinfo  {journal} {Nature Physics}\ ,\ \bibinfo {pages} {1}} (\bibinfo {year} {2024})}\BibitemShut {NoStop}%
\end{thebibliography}%
\end{document}